\documentclass[a4paper,onecolumn]{article}
\usepackage{blindtext}
\usepackage[a4paper, total={6.5in, 8.6in}]{geometry}
\usepackage{authblk}
\usepackage[english]{babel}
\usepackage{blindtext}
\usepackage{graphicx}
\usepackage{placeins}
\usepackage{float}

\title{Characteristics of partially filled Helmholtz resonators}
\author[1]{{Indenbom} M.V.}
\author[1,2,*]{{Pogossian} S.P.}
\affil[1]{Université de Bretagne Occidentale (UBO), 6, av. Le Gorgeu, 29238, 
	Brest, France}
\affil[2]{UBO Brest, CNRS, IRD, Ifremer, IUEM, Laboratoire d'Océanographie Physique et Spatiale (LOPS), \\
	Technopôle Brest-Iroise, Rue Dumont d'Urville, 29280, \\
	Plouzané, France}
\affil[*]{Corresponding author: S.P. Pogossian, pogossia@univ-brest.fr}

\date{17/12/2022}                     
\begin{document}
  \maketitle

	\begin{abstract}
	In this work we have developed a technique for the measurement of the resonance curve of Helmholtz resonators as a function of filling with beads and sands of different sizes, and water as the reference.  Our measurements allowed us to observe very different behaviors of resonance frequencies and resonance half-widths as a function of the size of the sand grains and the beads. 
	By comparing results for beads and water we were able to prove that the sound penetrates the interstitial space between the beads.  
	This was confirmed by measurements of the resonance properties in resonators filled with spherical beads whose experimentally determined filling factor is close to the random filling factor of about 0.54.
	The similar behavior of the frequency and half-width of resonance of sand-filled resonators of three different sizes allowed us to suggest that sound penetrates the sand in the same way as it does the beads.
	\end{abstract}


\section{\label{sec:1} Introduction}
The resonators invented by Helmholtz was originally intended for analyzing the tones of musical sound \cite{Helmholtz1954},\cite{Dosch2018}. Today the field of application of the acoustic Helmholtz resonator (HR) is very wide. The control of frequency properties such as the quality factor and the frequency of the resonance are important for a large number of applications in modern technology in various military and civil sectors. 
In architecture due to the concentration of population in urban areas, the use of absorbing structures to modify the architectural acoustics, the damping of low frequencies has proved to be a major objective \cite{Arenas2010}. 

To achieve the absorption of low frequencies due to noise pollution with the ongoing urban densification, textile resonators with multiple absorption mechanisms have been considered \cite{Neuwerk2020}. 

In oceanography, the concept of HR has been used to describe the dynamics of tidal elevation in a short, nearly enclosed basin that is connected to a tidal sea by a narrow strait \cite{Miles1975}, \cite{Doelman2002}. 

The HR  help to explain the functioning of earthen vessels in medieval buildings (also known as acoustic vessels) which have been of interest to archaeologists and acousticians \cite{Minns2020}.

It is well known that the resonance frequency shift can be used for measuring the volume of liquids in the resonator \cite{Nakano2006}, \cite{Webster2010}.  HR is also used as a measuring device, as well as in the fabrication of sensors \cite{Nishizu2001},\cite{Metcalfe2019}.  Belluci et al. \cite{Bellucci2004} analyzed the thermo-acoustic pulsations that occur in gas turbine combustion chambers and discussed the use of Helmholtz resonators to dampen the noise of these pulsations generated by rocket and aircraft engines. 

Car manufacturers use Helmholtz resonance for noise suppression which is one of the components of environmental noise pollution \cite{Yasuda2013}. Sometimes, new versatile materials like metamaterials are used for these purposes \cite{Hedayati2020}, \cite{Mahesh2019}, \cite{Yamamoto2018}.

\begin{figure}
	\centering
	\includegraphics[height=7.5 cm, width=5.14 cm,angle=0]{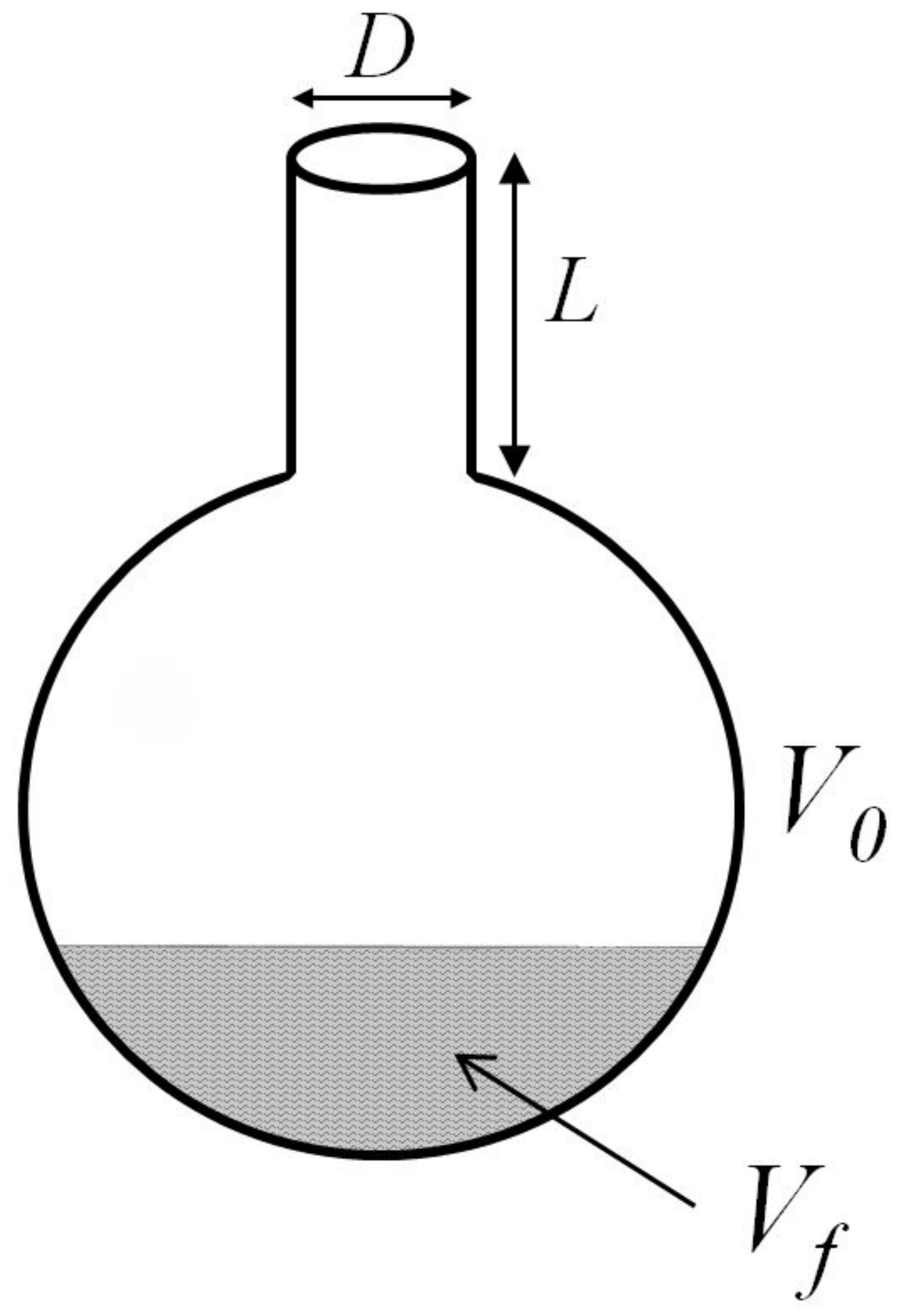}
	\caption{{Schema of a partially filled Helmholtz resonator with characteristic parameters \textit{$V_f, V_0, L$} and \textit{D}.}}
	\label{FIG1}	
\end{figure}

The main goal of our work is to study how partial fillings changes the resonance properties of  HR \cite{Nakano2006},\cite{Webster2010},\cite{Nishizu2001}. It is well known that the basic frequency \textit{$f_0$}   of the HR can be changed in a controlled manner by filling it partially for example with water Fig. \ref{FIG1}. 

The shift of the resonance frequency \textit{$f_0$} is due to the change of the resonator volume and can be described by the following equation \cite{Kinsler2000}, \cite{Morse1968}, \cite{Greene2009}:\\

\begin{equation}
	f_0=\frac{c_{air}}{2\pi}\sqrt{\frac{S}{V(L+0.6D)}}   
	\label{eq:1}
\end{equation} 
\\
Here $S = \frac{\pi D^2}{4}$ is the cross section area of the cylindrical neck of HR, \textit{D} is its diameter,  \textit{V = $V_f-V_0$} is the remaining free volume,  \textit{$V_0$} is the total volume of the empty resonator and \textit{$V_f$} is filled volume. The parameters of the resonator neck \cite{Chanaud1994},\cite{Mercier2017}: the cross section area \textit{S} and the length \textit {L}, as well as the sound speed \textit{$c_{air}$}   in air, do not depend on the filling. 

After a preliminary study of the influence of water filling on the resonance properties of HR as a reference, we extended our study to the variations of \textit{$f_0$} as a function of the filling level of the Helmholtz resonator by granular materials.

The paper is organized as following: In section \textit{Experimental Technique} we described our technique of the acoustic resonance measurements; in the \textit{Experimental Results and Discussion} section, measurements of the resonant frequency and linewidth are presented, as well as a discussion of the relevance and impact of our experimental results; finally in \textit{Conclusions} we outline the significance of the obtained results.

\section{Experimental technique\label{sec:2}}

In order to record the resonance curve the acoustic oscillations were excited by a small earphone put inside the HR. This technique excludes strong disturbing sound created when the resonance is excited by a loud speaker outside the resonator. The microphone registering the acoustic response is also placed inside the resonator.

\begin{figure}
	\centering
	\includegraphics[height=5.64 cm, width=8.5 cm,angle=0]{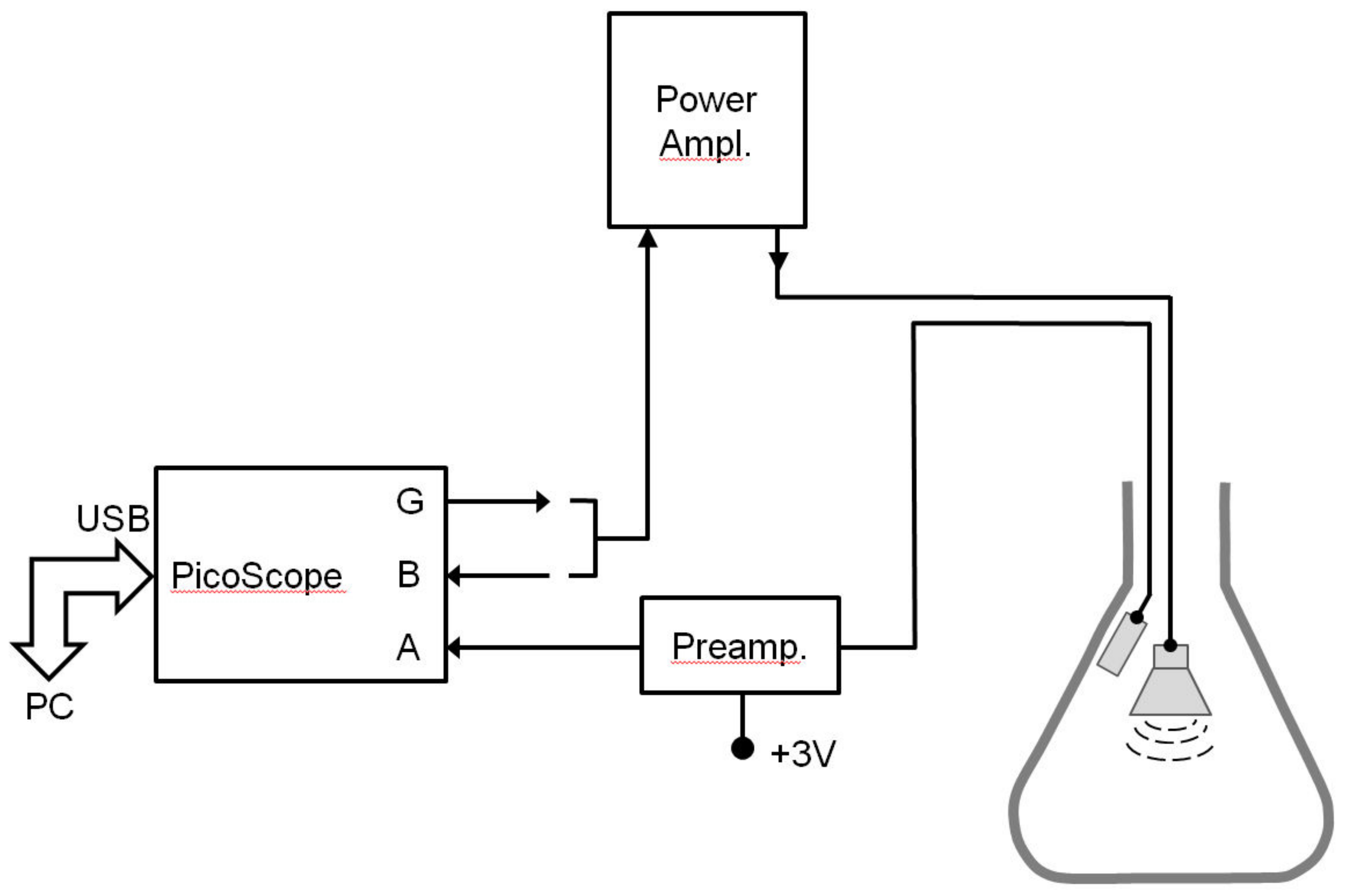}
	\caption{Schema of the measurements using the PicoScope.}
	\label{FIG2}
\end{figure}

Precise resonance curves were obtained using a small oscilloscope without screen, PicoScope 2204A, which was controlled by a PC  using Python. The PicoScope has an intrinsic signal generator which excites the earphone Fig. \ref{FIG2}. A power amplifier was used because of too high intrinsic resistance (600 Ohms) of the generator. The microphone is connected to the input A of the PicoScope via a preamplifier based on 2N-2222A triode. Its signal transmitted to PC is fitted by a sinusoidal curve having the generator frequency, so the exact amplitude of the acoustic signal is obtained even if the original signal is noisy. 
In addition to the amplitude, the acoustic phase could also be obtained by connecting the generator  output G to the PicoScope input B the signal of which is fitted by a sinusoidal curve. 

A highly accessible modification of the technique can be realized using a smartphone with phyphox application \cite{Staacks2018}, \cite{Stampfer2020} keeping practically the same measurement precision. The only drawback is the loss of phase information which is not used in the present work. A standard earphone with a microphone is connected directly to the smartphone and put inside the resonator (Fig. \ref{FIG2}). The frequency sweep and acoustic response are obtained using an appropriate phyphox experiment. The obtained curves are saved and replotted using a Phython code.

For all measurements in this paper a 100 $cm^3$ Erlenmeyer flask (conical flusk) was used as the HR. The internal volume of the resonator \textit{$V_0$}  is about 115 $cm^3$. The neck has a length \textit{L}  =  32 mm and a diameter \textit{D}  = 18 mm. 

\begin{figure}
	\centering
	\includegraphics[height=5.8 cm, width=8.5 cm,angle=0]{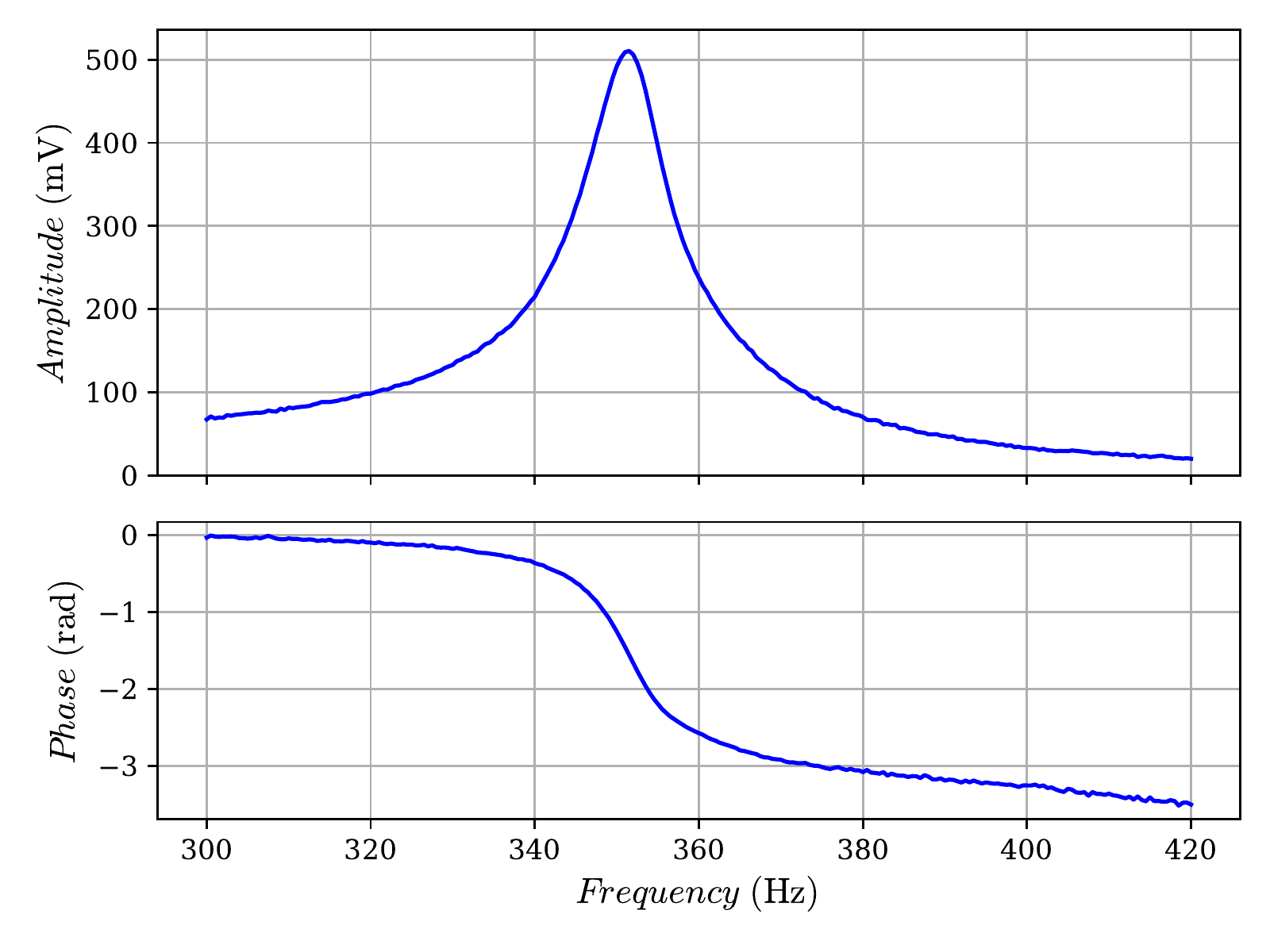}
	\caption{The resonance behavior of the acoustic amplitude and the phase in function of the excitation frequency demonstrates a close analogy of the Helmholtz resonator and a classical spring-mass oscillator.}
	\label{FIG3}
\end{figure}

The resonance frequency estimated by Eq. \ref{eq:1} is about 391 Hz (here the value of the sound speed in air $c_{air}$  = 343.5 m/s at 20$^{\circ}$C is used). Measurements using PicoScope give a somewhat lower value 351 Hz (see Fig. \ref{FIG3}). The difference is due to a deviation from assumptions \textit{$SL/V\ll1$}  and \textit{$S/V^{2/3} \ll 1$} of Eq. \ref{eq:1} derivation (here \textit{$SL/V$} = 0.04 and \textit{$S/V^{2/3}$} = 0.11).

The acoustic resonance is very similar to the classical resonance of a mass on a spring. The resonance peak shape corresponds to one of the mass displacement and the difference of the phase of acoustic pressure oscillations follows classical behavior: at low frequencies the excitation and the response are in-phase, at the resonance frequency the phase shift is 90$^{\circ}$ and at high frequencies the excitation and the response are anti-phase (Fig. \ref{FIG3}).  

Physically compressing–decompressing of air in the volume of the HR accompanied by air  motion in the neck is an analogue of the mass-spring mechanical oscillator system: the spring corresponds to the volume of the gas in the resonator and the mechanical mass to the mass of moving air in the neck \cite{Kinsler2000}.

\section{Experimental results and discussions\label{sec:3}}

Before studying the variation of the resonator frequency as a function of the filling level with granular materials, we first measured the frequency behavior of the resonator as a function of the partial filling with water as a reference.

\begin{figure}
	\centering
	\includegraphics[height=5.8 cm, width=8.5 cm,angle=0]{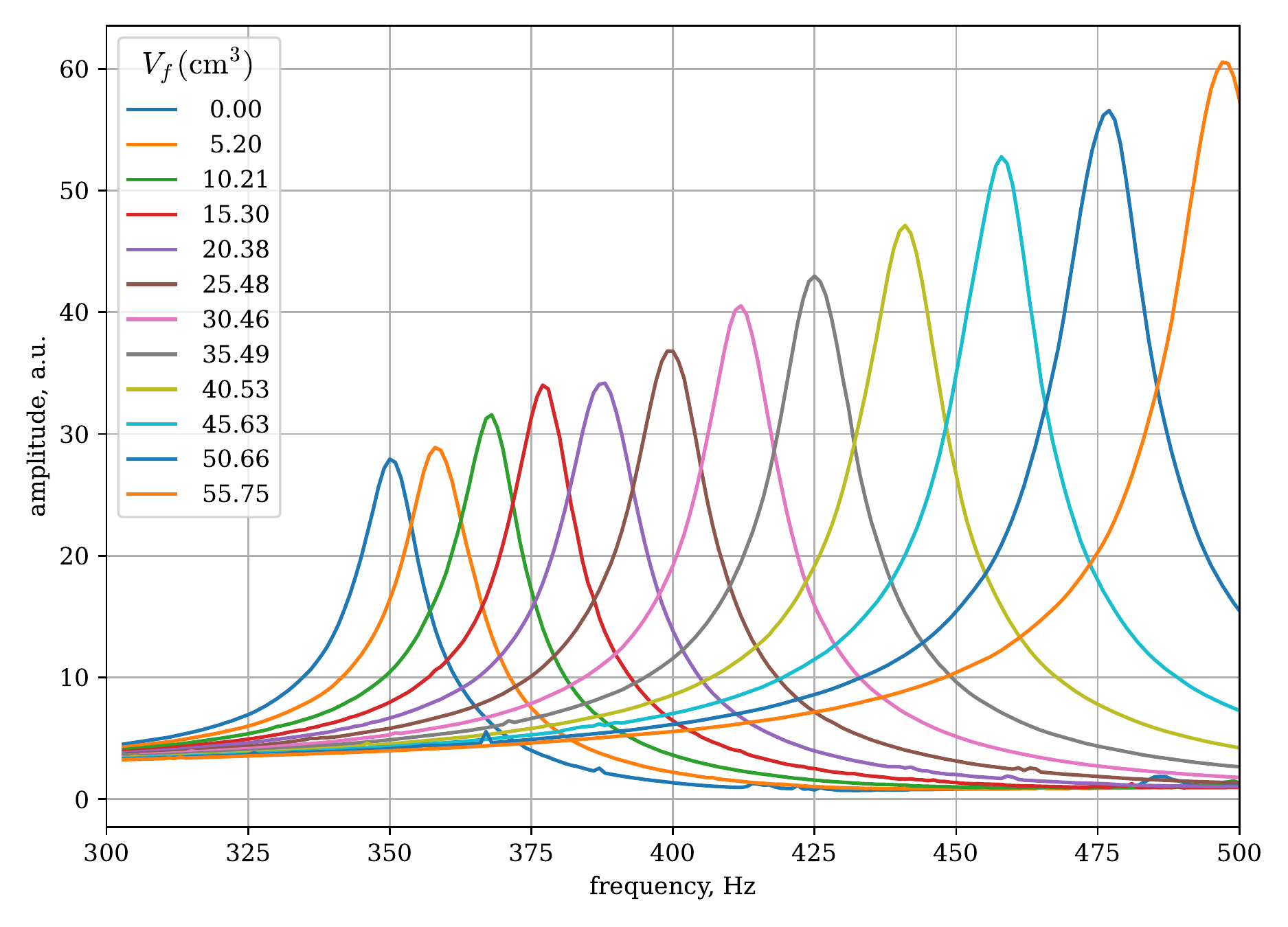}
	\caption{Modification of the acoustic resonance by filling the Helmholtz resonator with water as known for any liquid \cite{Nakano2006}, \cite{Webster2010}.}
	\label{FIG4}
\end{figure}

The resonance peak of our Erlenmeyer flask is gradually shifted to high frequencies upon filling with water (Fig. \ref{FIG4}). Resonance peak is shifted from the left to the right with increasing the water volume \textit{$V_f$} by step of about 5 $cm^3$. The values of resonance frequencies were refined by fitting of 5 points at the top of each resonance peak by a parabola. 

\begin{figure}
	\centering
	\includegraphics[height=5.8 cm, width=8.5 cm,angle=0]{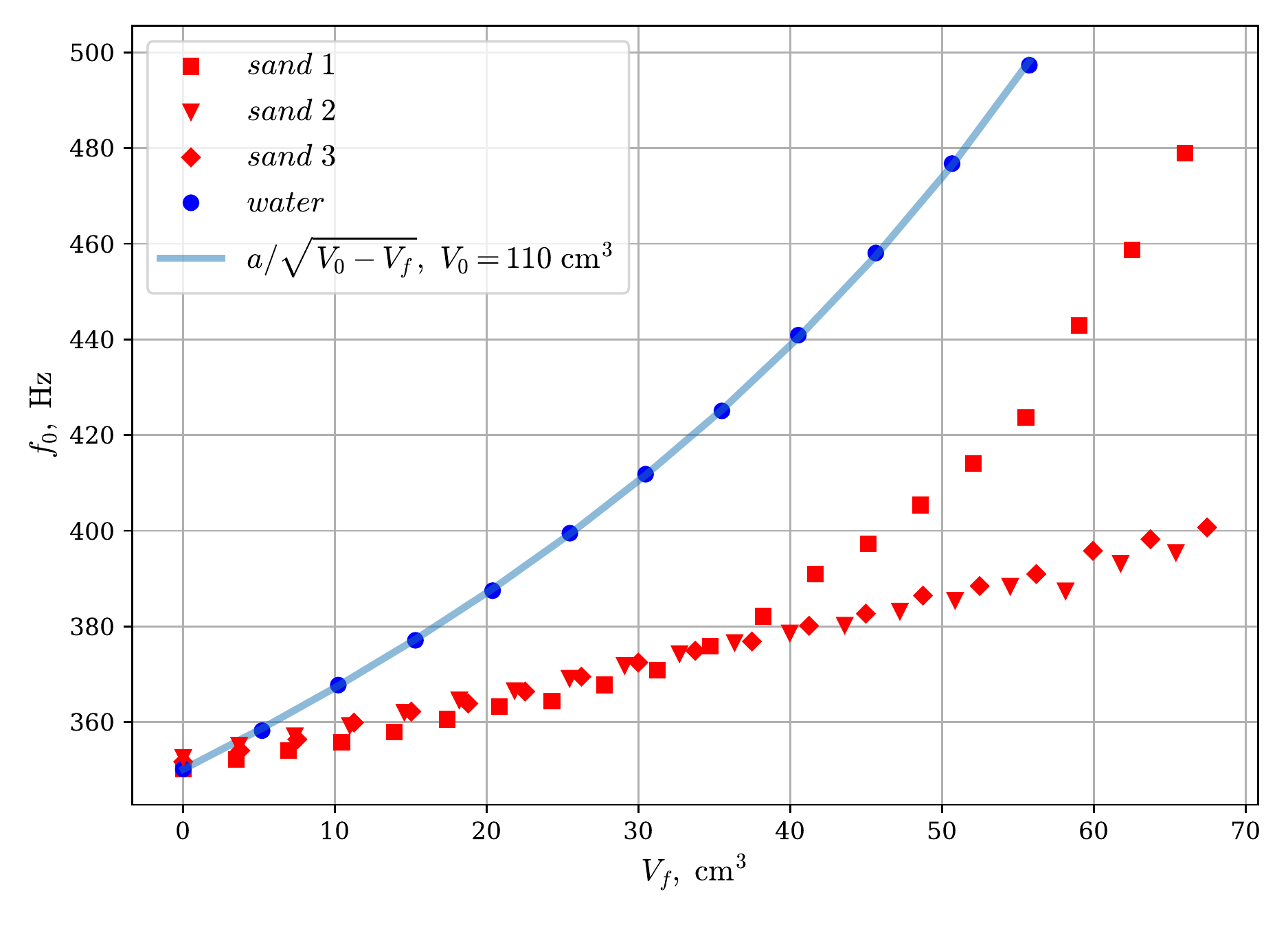}
	\caption{Resonance frequency \textit{$f_0$} dependence on filled volume \textit{$V_f$} for water (circles), sand 1 (squares) sand 2  (triangles) and sand 3 (by diamonds).}
	\label{FIG5}
\end{figure}

Sound practically does not penetrate into water and the resonator acting volume \textit{V}  is the difference between its initial volume \textit{$V_0$} and the water volume \textit{$V_f$}. It is proved by a fitting of the experimental points by the function $f_0=\frac{a}{\sqrt{V_0-V_f}}$ with two fitting parameters \textit{a}  and \textit{$V_0$}  (Fig. \ref{FIG5}), as seen from Eq. \ref{eq:1}. Here we obtain \textit{$V_0$} =110 $cm^3$ that is very close to our measured value of 115 $cm^3$. 

\begin{figure}
	\centering
	\includegraphics[height=16.0 cm, width=7.1 cm,angle=0]{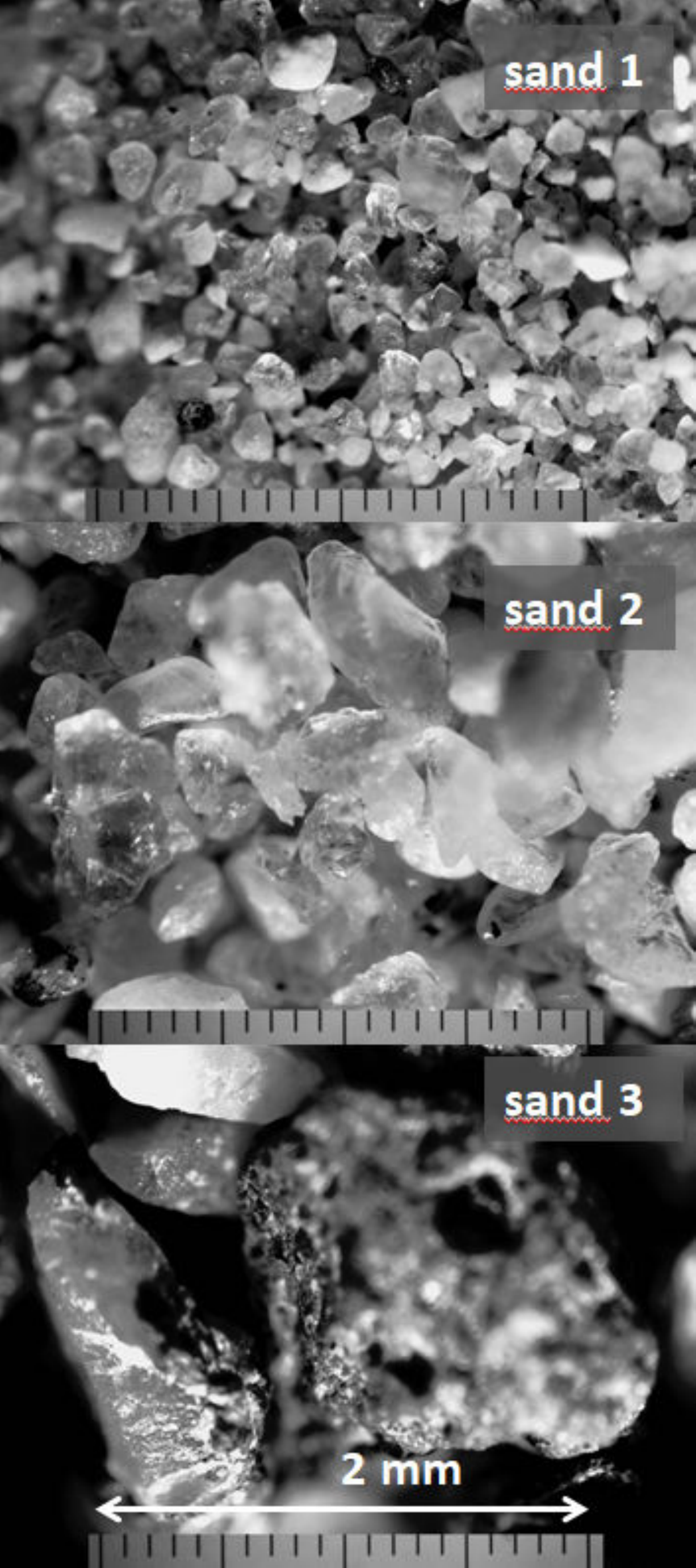}
	\caption{Photos of three sands used in this paper have the same scale.}
	\label{FIG6}
\end{figure}

Different results were obtained in the following measurements of resonance frequency as a function of the resonator partial filling with three different sands that we will call sand 1, sand 2 and sand 3 having respectively grain sizes $ 0.2 \pm 0.1 $ mm, $0.4 \pm 0.2$ mm and about $ 1.1 \pm 0.5 $  mm. Fig. \ref{FIG6} shows microphotos of these sands at the same scale. The formation of sand grains by the decomposition of pre-existing rocks by weathering and erosion results in a significant fluctuation in the size of these grains \cite{Nichols2009}. 

The resonance curves of these sands have the same frequency evolution as a function of the filling volume up to 30 $cm^3$ as seen on Fig. \ref{FIG5}. From this frequency, the evolution of sand 1 changes its quasi linear behavior and becomes nonlinear similar to that of water. For comparison reasons we see on Fig. \ref{FIG5} the frequency behavior of water as a function of the filling volume which follows perfectly the theoretical predictions described \cite{Kinsler2000}, \cite{Greene2009} by Eq. \ref{eq:1} .  Note that the samples sands 2 and 3 continue to have a quasi-linear behavior up to the maximum filling volumes of 70 $cm^3$. 

\begin{figure}
	\centering
	\includegraphics[height=6 cm, width=8.5 cm,angle=0]{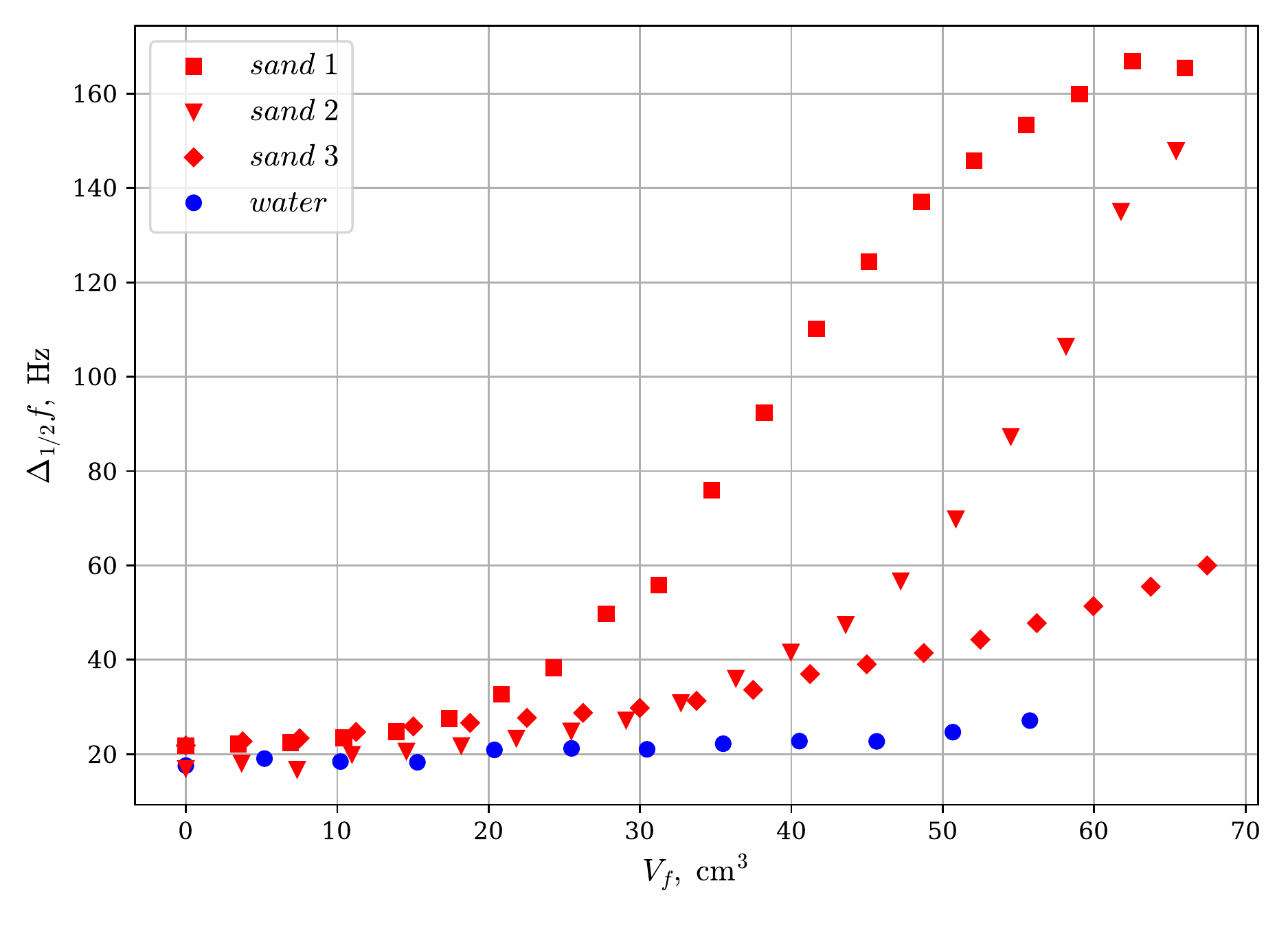}
	\caption{Resonance half-width $\Delta_{1/2}f$ dependence on filled volume \textit{$V_f$} for water (circles), sand 1 (squares) sand 2  (triangles) and sand 3 (by diamonds).}
	\label{FIG7}
\end{figure}

These differences in behavior between sands 2,3 and sand 1 are even more pronounced in the trends of the resonance half-widths as a function of the filling volume as shown in Fig. \ref{FIG7}.

The measurements of the half-widths of resonance as a function of the volume of filling demonstrate clear difference between the dependencies of the of the water and the 3 sands.  The half-widths of three sands of different sizes increase gradually up to value  \textit{$V_f$} =20 $cm^3$ with a slope about the same than that of the water. Then a distinct upturn  occurs for the sand 1 with the fastest growth that comes to a saturation at about \textit{$V_f$} = 60 $cm^3$.   Sand 2 follows the same type of deviation from the quasi-linear behavior of water with an upturn at \textit{$V_f$} = 35 $cm^3$.  The behavior of sand 3  is clearly different from that of water but has a slower evolution than sand 1 and sand 2. The significant deviation of the behavior of the resonance frequency and the resonance half-width of three sands as a function of \textit{$V_f$} from that of water makes us believe that there is a penetration of sound into the interstitial volume between the sand grains.

To better understand these differences in the behavior of three sands, and to be able to assert the hypothesis of the penetration of sound in the sand we decided to study two types of spherical beads of different well defined diameters and different densities.  Obviously,  the packing factor has a great importance in this problem. The packing factors of the sands are difficult to evaluate since they contain grains of different sizes and of very irregular shapes. For large diameter spherical beads (d=5.92 mm), the packing factor can be exactly determined experimentally since they can be easily counted.

\begin{figure}
	\centering
	\includegraphics[height=6 cm, width=8.5 cm, angle=0]{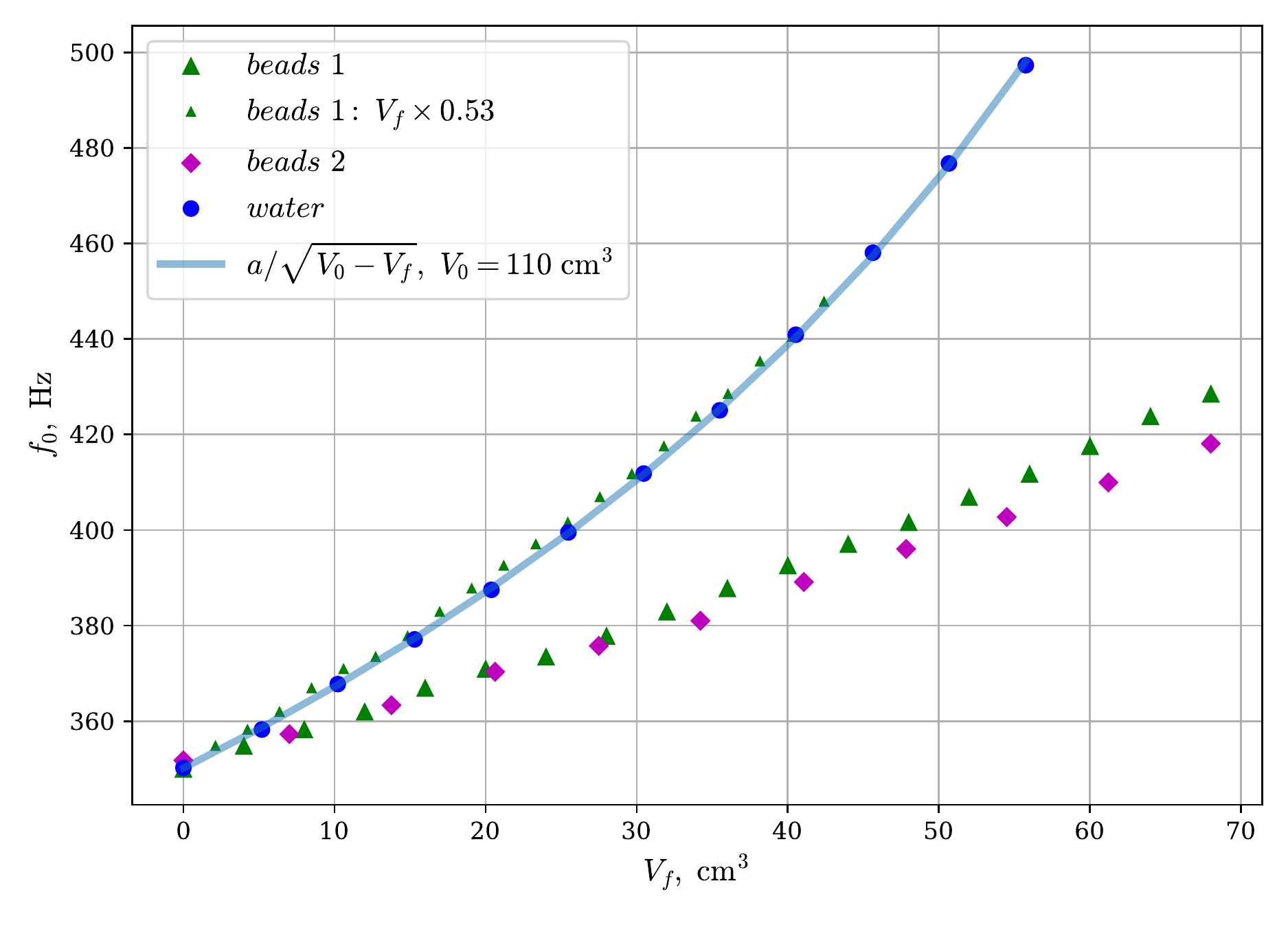}
	\caption{Resonance frequency $f_0$ dependence on filled volume  for water (circles), beads 1 (big triangles),  beads 2  ( diamonds). Small triangles show the data for the beads renormalized to their proper volume \textit{$V_f\times$}0.53 (the renormalization coefficient 0.53 corresponds to the loose random packing factor of spheres).}
	\label{FIG8}
\end{figure}

For this purpose, we have compared in Fig. \ref{FIG8} the behavior of the resonant frequency of the resonator filled with water and with spherical beads of 2 different sizes. 
From now on, we will call beads 1, the ones of diameter $d_1=5.92 \pm 0.01 mm$ and, beads 2 the ones of diameter $d_2=0.9 \pm 0.1$ mm.

As can be seen in Fig. \ref{FIG8}, beads 1 and 2 of different sizes have similar behavior. The evolution of the frequency as a function of \textit{$V_f$}  for beads 2 deviates slightly more from the behavior of water.  
We must distinguish 2 volumes that can be measured experimentally. The volume \textit{$V_f$}  designates the total volume occupied by all the beads (1, or 2). 

For example 375 balls of diameter 5.92 mm occupy a volume of \textit{$V_f$}=76 $cm^3$ in the experimental test tube. Each spherical bead has its own volume. By adding up the volumes of all the beads, one gets the proper volume that all the beads have together. This volume differs from \textit{$V_f$} by a coefficient that is called the packing factor, which characterizes the density and compactness of the packing of these spheres. Taking into account that the volume of a single bead is 0.109 $cm^3$, the 375 beads have a total proper volume of 40.7 $cm^3$. The packing factor will then be equal to ratio of these two volumes 0.536 = 40.7$/$76. The way that the beads were introduced into the HR leads us to believe that the resulting packing can be considered as loose random packing. The loose random packing factor is well known and has been was calculated and observed experimentally for single size spherical beads in a number of works \cite{Panayiotopoulos1989}, \cite{Jefferies1993},  \cite{Song2008}, \cite{Farrell2010}, \cite{Torquato2010}.

Taking into account the loose random packing factor and correcting the filling volume by this factor we see that the behavior of the resonance frequency as a function of the proper volume of all beads is almost identical with that of water.  

\begin{figure}
	\centering
	\includegraphics[height=6 cm, width=8.5 cm,angle=0]{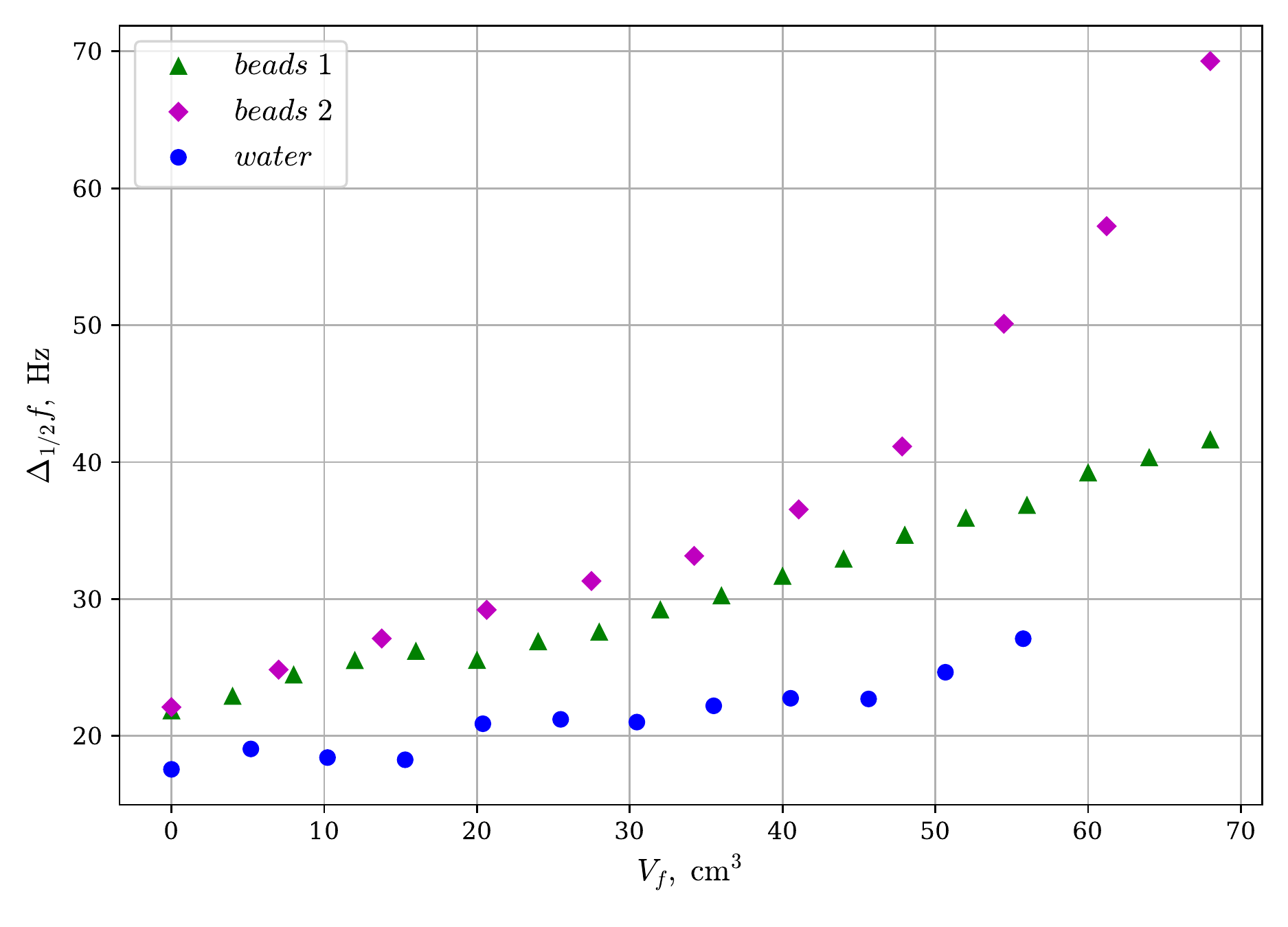}
	\caption{Resonance half-width $\Delta_{1/2}f$  dependence on filled volume \textit{$V_f$} for water (circles), beads 1 ( triangle),  beads 2  ( diamonds).}
	\label{FIG9}
\end{figure}

On the Fig. \ref{FIG9} we see that the half-widths of resonance of the two beads of different sizes increase more rapidly than the half-width of resonance of the water.  The evolution of half-width of resonance of the water and of big beads (beads 1) as a function of \textit{$V_f$} are quasi linear. The curve of half-width of resonances of the small beads 2 is superior to that of the water and beads 1 and undergoes a fast growth starting from \textit{$V_f$}  = 40 $cm^3$, which   approaches the behavior of sand 2 and 3 (see Fig. \ref{FIG7})  remaining between these two curves. It is worth noting that the size of the beads 2 is larger than that of the sand 2 and smaller than that of the sand  3. The fact that the half-widths of resonance of the 2 beads is above the water curve is explained by the fact that there is damping of the sound in the interstitial channels between the beads. This statement is justified by the observation that the damping is greater for small beads because the corresponding interstitial channels are thinner \cite{Johnson1987}, \cite{Leclaire1996}, \cite{Allard2009}.

\section{Conclusions\label{sec:4}}
Based on our measurement technique we have shown in this work that the study of the resonance half-width is more sensitive to the processes of sound penetration into the filling substance and damping of sound oscillations than the corresponding behavior of the resonance peak. 

Our observations of the frequency and the half-width of the resonance peak as a function of water filling are consistent with well-known results and confirm the non-penetration of sound oscillations in the latter.  

The properties of the resonance with sand filling are much less known and understood. The increase of the resonance frequency as a function of the amount of sand is much less important than that of water and initially has an almost linear course. More interestingly, for sand with small grains (sand 1) a rapid increase of the resonance frequency following the behavior of water is observed. The sands of larger size continue to increase with their initial trend without notable differences. 

More varied trends are observed on the resonance half-width curves. The curves of three sands with three different sizes are distinct. The sand with the smallest grains shows an upturn in the curve at low fillings. We observe this tendency for the 2 other sands with the upturns depending on their size.

One observes with the beads of different size also weaker increase of resonance frequency compared to water.  The measurement of the packing factor of the beads and the comparison with the resonance behavior of water shows that this small increase in resonance frequency of the beads is related to the total penetration of the sound in the interstitial space between the beads. This penetration becomes even more obvious in the behavior of the resonance half-width as a function of partial filling. For smaller beads, narrow interstitial channels favor the increase of viscous damping and therefore resonance half-width. 

This analysis shows that a similar sound penetration into the interstitial space in the sands is responsible for both the small increase in resonant frequency related to the packing factor and for the very specific behavior of the resonance half-width curves controlled by the viscous damping.

\section*{Acknowledgement}
Authors thank Olivier Roux for technical assistance


\end{document}